\documentclass[pdflatex,sn-basic]{sn-jnl}

\makeatletter
\renewcommand{\NAT@aysep}{, }
\makeatother

\usepackage{graphicx}%
\usepackage{multirow}%
\usepackage{amsmath,amssymb,amsfonts}%
\usepackage{amsthm}%
\usepackage{mathrsfs}%
\usepackage[title]{appendix}%
\usepackage{xcolor}%
\usepackage{textcomp}%
\usepackage{manyfoot}%
\usepackage{booktabs}%
\usepackage{algorithm}%
\usepackage{algorithmicx}%
\usepackage{algpseudocode}%
\usepackage{listings}%
\usepackage{amssymb}
\usepackage{pifont}
\newcommand{\cmark}{\ding{51}}

\begin{document}

\title[The Decentralisation Paradox in Digital Identity: Centralising Decentralisation with Digital Wallets?]{The Decentralisation Paradox in Digital Identity: Centralising Decentralisation with Digital Wallets?}

\author[1]{\fnm{Ioannis} \sur{Konstantinidis}}

\author[1]{\fnm{Ioannis} \sur{Mavridis}}

\author[2]{\fnm{Evangelos K.} \sur{Markakis}}

\affil[1]{\orgdiv{Department of Applied Informatics}, \orgname{University of Macedonia}, \country{Greece}}

\affil[2]{\orgdiv{Department of Electrical and Computer Engineering}, \orgname{Hellenic Mediterranean University}, \country{Greece}}

\abstract{Digital identity is shifting from service- and network-centric approaches toward user-centric ones that promise users increased control over their data. Despite their decentralised design, such approaches often reintroduce centralised components in different forms. This paper conceptualises this tension as the decentralisation paradox and shows that user-centric architectures tend to redistribute rather than eliminate centralisation. Drawing on Critical Systems Thinking (CST), digital identity is framed as a "wicked problem" that spans the technical, legal, social and ethical dimensions. It introduces the Digital Identity Tetrahedron as a multidimensional framework for analysing decentralisation in digital identity ecosystems. Understanding these interdependencies is essential for designing reliable architectures and ensuring that the next generation of digital identity surpasses superficial decentralisation.}


\maketitle

\section{Introduction}
Who owns digital identities? In user-centric digital identity approaches, including decentralised identity and Self-Sovereign Identity (SSI), users hold and control their credentials in digital wallets and decide when and with whom to share them \citep{Babel-2025, Cucko-2022, Sedlmeir-2021}. However, research indicates an ongoing tension between between decentralisation and the continued reliance on centralised forms of co-ordination, governance and control \citep{Degen-2024, Tan-2023}, which this paper conceptualises as the “decentralisation paradox”. As a consequence, rather than eliminating centralisation or reducing it to a minimum, user-centric approaches often redistribute it across the broader ecosystem.

Despite the promise of user-centric approaches in fostering innovation and re-enforcing privacy and data protection, they remain constrained by a number of co-ordination-related challenges \citep{Krul-2024, Tan-2023}, as well as fragmented standards \citep{Lips-2022, Supangkat-2025, Yildiz-2023}. The literature suggests that decentralisation operates along a continuum, where architectures are decentralised in some aspects and centralised in others \citep{Cucko-2022, Manimaran-2025, Satybaldy-2024, Yildiz-2023}. In particular, governmental entities and institutional actors often reassert a central position even in architectures that incorporate decentralised components \citep{Alizadeh-2022, Gans-2022, Krul-2024}. These observations indicate that decentralisation—in digital identity—is characterised by systemic dependencies and the redistribution of control, trust and governance across multiple actors within ecosystems.

While the academic discourse identifies specific manifestations of the decentralisation paradox, the existing contributions typically analyse decentralisation from a single perspective (e.g., architectural design or legal considerations). To date, the decentralisation paradox has not been systematically examined as a multi-dimensional phenomenon. This paper addresses this gap by providing a holistic analysis of the decentralisation paradox in digital identity, treating it as an emergent property of a broader socio-technical system rather than a purely architectural characteristic.

Given the inherent complexities of modern digital identity, including multi-level socio-technical interactions, cross-jurisdictional recognition and adoption dynamics \citep{PavaDiaz-2024}, this paper adopts Critical Systems Thinking (CST) \citep{Jackson-2019} as its methodological approach. CST enables a holistic examination of socio-technical systems by analysing interactions across multiple dimensions and highlighting how interventions in one dimension might introduce unintended consequences in the others. Through CST, digital identity is conceptualised as a “wicked problem” \citep{Jackson-2024} that involves interdependent technical, legal, social and ethical considerations and resists simple solutions. As such, the exploration of this decentralisation paradox leads to the following Research Questions (RQs):

\begin{enumerate}
\item[] \textbf{RQ1:} How does digital identity evolve from service-centric and network-centric to user-centric approaches, and how does this evolution contribute to the emergence of the decentralisation paradox?

\item[] \textbf{RQ2:} How do the technical, legal, social, and ethical dimensions interact to create the decentralisation paradox in user-centric digital identity ecosystems?

\item[] \textbf{RQ3:} How do digital wallets and their surrounding architectural components operate as new intermediaries in user-centric digital identity ecosystems, and how does this reinforce the decentralisation paradox?
\end{enumerate}

The contribution of this research is both analytical and conceptual. First, it systematically articulates the decentralisation paradox in user-centric digital identity, demonstrating that decentralised architectures redistribute, rather than eliminate, centralisation. Second, it identifies how additional intermediaries emerge within user-centric ecosystems, thereby reintroducing co-ordination mechanisms even in decentralised architectures. Third, the paper proposes the “Digital Identity Tetrahedron” as a structured framework for analysing centralisation dynamics across the technical, legal, social and ethical dimensions. Together, these contributions advance the theoretical understanding of user-centric digital identity, with the intention of supporting more context-aware design choices and informed policy-making decisions.

This research is subject to certain limitations. The conceptual nature of the analysis does not provide an empirical validation of the proposed framework and the interpretations of decentralisation might differ across jurisdictional contexts. Furthermore, the Digital Identity Tetrahedron focuses on the technical, legal, social and ethical dimensions, while additional factors—such as commercial or organisational considerations—might also influence the decentralisation dynamics. These aspects provide opportunities for future research and empirical examination.

\section{Evolution of Digital Identity (RQ1)}

\subsection{Service-, Network- and User-Centric Approaches}
Digital identities are understood as sets of attributes that represent entities within a specific context \citep{ISOIEC-24760-1-2025, ITU-T-X1252-2021, NIST-2024-SP800-63}. They have evolved from service-centric to network-centric approaches, and more recently towards user-centric ones that often incorporate the concepts of decentralised identity or SSI \citep{Alizadeh-2022, Manimaran-2025, Schumm-2025}. In this paper, user-centric digital identity is treated as a superset that encompasses decentralised and SSI-aligned architectures, where digital wallets enable users to store credentials and manage the disclosure of identity data \citep{Konstantinidis-2026}. This evolution reflects a broader paradigm shift in how digital identities are created, managed and used across ecosystems.

In service-centric approaches, digital identities are managed in a centralised manner, where a single Identity Provider (IdP) issues credentials and manages data within a specific environment. These approaches suffer from a number of limitations \citep{Schardong-2022, Soltani-2021}, such as credential proliferation and password fatigue. In response, network-centric approaches, including Federated Identity Management (FIM) and Single Sign-On (SSO), establish trust relationships among IdPs and Service Providers (SPs) and enable users to access multiple SPs with the same set of credentials. While these network-centric approaches offer a consistent experience and ease of use, they still revolve around a limited number of IdPs \citep{Manimaran-2025, Soltani-2021}. As a consequence, trust and control remain centralised, often rendering these architectures single points of failure.

\begin{figure}[!t]
\centering
\includegraphics[width=3.1in]{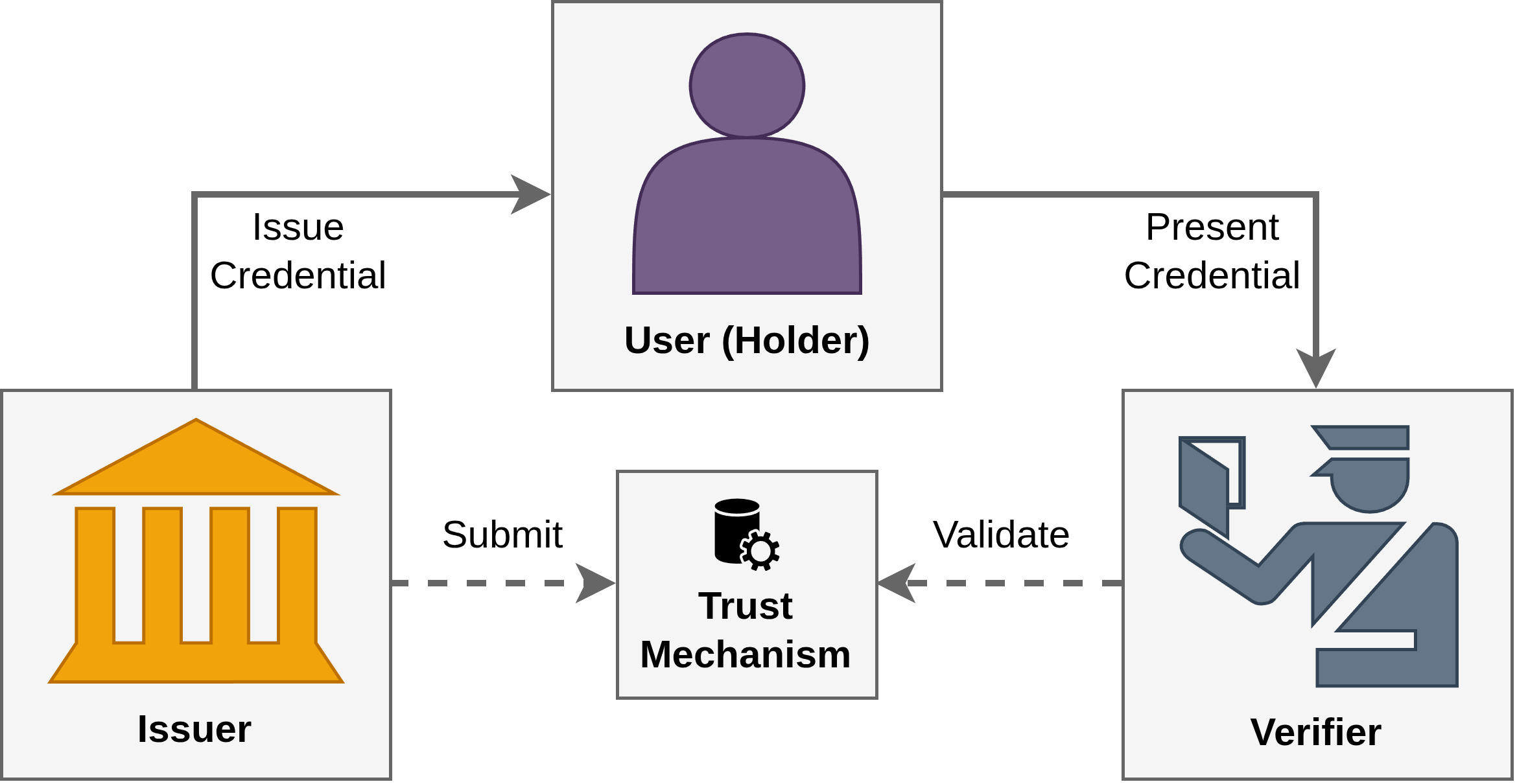}
\caption{An illustration of a user-centric architecture that depicts the issuer, user, verifier and the underlying trust mechanism. The trust mechanism enables the validation of credentials without direct interactions between the issuers and verifiers.}
\label{fig:user-centric-architecture}
\end{figure}

With the emergence of Distributed Ledger Technology (DLT), user-centric approaches began supporting users in managing their own credentials through cryptographic mechanisms and blockchain-based registries \citep{Alizadeh-2022, Cucko-2021, Schumm-2025}. A cornerstone of these approaches are digital wallets, i.e., solutions that hold users’ credentials and cryptographic keys. Figure \ref{fig:user-centric-architecture} indicates the three main actors in such architectures: a) issuers that issue credentials to users, b) users who obtain and store credentials in their digital wallets and c) verifiers that make decisions based on the presented credentials \citep{Cucko-2022, Babel-2025}. As credentials are portable and under the users' possession, it is possible to present one credential across multiple SPs or combine data into a single presentation.

It is often argued that decentralisation eliminates the before-mentioned single points of failure. Moreover, the reduction of centralised IdPs could mitigate the risk of mass data breaches and user surveillance \citep{Krul-2024, Manimaran-2025, Schumm-2025}. However, these promises face significant practical challenges, as user-centric implementations are not necessarily based on DLT but often depend on centralised components such as Public Key Infrastructure (PKI) \citep{Konstantinidis-2026}. Even when standardised protocols exist, the deployment of such user-centric approaches at a national or international scale is subject to multiple technical, legal, social and ethical considerations \citep{Degen-2024, Ishmaev-2021, Krul-2024, Lockl-2023, Manimaran-2025, Martin-2024, Satybaldy-2024, Schardong-2022, Schumm-2025, Sharif-2025, Pohn-2024, Ziegler-2025}. It is not sufficient to develop a digital wallet or a credential verification mechanism; one must also establish frameworks, unified standards and adoption strategies for an architecture to operate in practice \citep{Cucko-2022, Soltani-2021, Weigl-2023}. This is where the paradox becomes evident: to replace IdPs, as seen in service- and network-centric approaches, additional forms of agreements and technical components are required.

\subsection{Trust and Co-ordination in User-Centric Approaches}
Both trust and co-ordination are foundational elements of digital identity ecosystems. When an SP relies upon a credential, there is an implicit trust that it is valid, it belongs to the legitimate user and the issuer of that credential is authoritative \citep{Krul-2024, RamosFernandez-2024}. In service-centric or network-centric approaches, establishing trust is straightforward, since the IdP serves as the trust anchor. Whereas, in user-centric approaches, trust is established through a combination of technical components (e.g., cryptographic mechanisms) and governance frameworks (e.g., policies, certification and oversight mechanisms) \citep{Gans-2022, Giannopoulou-2021}. In fact, the concept of a “trust framework” is often used to denote a set of agreements, standards and roles that enable trust across contexts and jurisdictions.

In user-centric approaches, trust frameworks take different forms, including the notion of “trustlessness” \citep{Cucko-2021, Krul-2024} which differs from PKI-based approaches. In this sense, trustlessness aims to reduce reliance on intermediaries; however, technical trustlessness still depends on governance structures such as trust registries \citep{Cucko-2023}. One needs to trust that issuers are authoritative and issue genuine credentials and that the overall framework is not compromised. In real-world implementations, what is often described as trustlessness is better characterised as "trust minimisation" \citep{Konstantinidis-2026}.

The literature discusses two main categories of trust-establishment mechanisms:
\begin{itemize}
  \item Web of Trust \citep{Ghaffari-2022, Soltani-2021}: Each verifier decides which issuers to trust in an independent manner, often through bilateral or community-based relationships. While decentralised, this approach becomes impractical at a large scale, as it lacks a common baseline of trust and could lead to fragmentation.
  \item Trust Anchors \citep{Cucko-2022, Krul-2024}: Trust registries co-ordinate and formalise trust relationships. In practice, these often require foundations, consortia or governmental entities to establish rules (e.g., requirements for issuers) and maintain lists of trusted issuers. Although credentials are still exchanged in a peer-to-peer manner, this approach reintroduces centralised points of co-ordination.
\end{itemize}

In this context, assurance is an important requirement \citep{Andrasko-2021, Schardong-2022} to take into consideration. For example, the eIDAS Regulation defines three Levels of Assurance (LoAs), i.e., Low, Substantial or High \citep{EU-2015-1502}. Likewise, NIST SP 800-63-3  \citep{NIST-2024-SP800-63A} specifies three Identity Assurance Levels (IALs). In essence, a higher LoA or IAL implies stringent identity proofing and verification processes such as in-person appointments and/or biometric recognition. This is where decentralisation once again introduces a pressure for centralisation.

In practice, such assurance and governance-related requirements also lead to the establishment of certification schemes. These are centralised, even if implemented through a network of auditors or authorities \citep{ARF-2025-Certification, RamosFernandez-2024}. The paradoxical outcome is the re-distribution of trust, i.e., users do not depend on a single IdP but still depend on a centralised governance mechanism \citep{Cucko-2022, Ghaffari-2022}. As the next section outlines, the real-world implementations of user-centric approaches at scale reveal important contradictions.

\subsection{Real-World Implementations of User-Centric Approaches}
A number of national and international implementations of user-centric approaches draw on principles from decentralised identity and SSI. These approaches position users as holders of credentials in digital wallets, with control over the disclosure of their identity data. In practice, however, several state-backed and government-led implementations incorporate centralised elements for trust, co-ordination, assurance and governance; this again reflects the broader decentralisation paradox.

In the European Union, the European Digital Identity Wallet (EUDI Wallet) is a prominent example under the eIDAS 2.0 Regulation \citep{EU-2024-1183}. Its architecture incorporates a number of open standards such as IETF SD-JWT VC, ISO/IEC 18013-5, ISO/IEC 23220-2, OID4VP and OID4VCI \citep{ARF-2025-Ecosystem}. In particular, the EUDI Wallet operates on a PKI-based trust mechanism \citep{ARF-2025-TrustModel} and each member-state maintains trusted lists of entities. Although this arrangement contributes to both assurance and interoperability, it also re-introduces centralised components. Furthermore, the proposed design raises concerns about unlinkability and observability \citep{Alvarez-2026, Baum-2024}. In PKI-based architectures, there are often observable co-ordination points that generate auxiliary metadata; in turn, these could be correlated across verifiers. As a result, even when selective disclosure is theoretically supported, the surrounding infrastructural components could still enable indirect linkability. In effect, the EUDI Wallet reflects an implementation where the appearance of user control coexists with strong institutional oversight, raising critical questions about the actual degree of decentralisation of such architectures.

In the United States, mobile Driving Licences (mDLs) are available in several states \citep{Arizona-mDL, California-mDL, Georgia-mDL}. These are provisioned to smartphone wallets and signed by the issuing authorities. While the interface is user-centric, the underlying trust mechanisms remain centralised. Furthermore, in certain implementations, there are additional dependencies: Apple Wallet \citep{Apple-Wallet-ID-States} and Google Wallet \citep{GoogleWallet-2025} support mDLs but introduce a degree of proprietary integration into a digital public infrastructure. Despite the promise of user control, the architecture revolves around centralised issuance, verification and governance as seen in network-centric approaches. What emerges is a user-centric interface on top of an institutional framework which raises questions about the degree of decentralisation and the practical limits of user control in recentralised architectures.

In Bhutan, the National Digital Identity (NDI) supports a range of credentials such as national identification documents, permits and certificates \citep{UNDP-Bhutan-DigitalStrategy-2024, ToIP-BhutanNDI-CaseStudy-2024} using W3C DIDs and W3C VCs. In principle, the NDI also supports selective disclosure and aligns with SSI in terms of user control and data minimisation. Its trust anchor is DLT-based, evolving from Hyperledger Indy to Polygon and later to Ethereum. Nevertheless, even the NDI reflects the decentralisation paradox, as it relies on authoritative trust registries and revocation mechanisms despite its decentralised infrastructure \citep{ToIP-BhutanNDI-CaseStudy-2024}. A national legal framework, i.e., the NDI Act \citep{Bhutan-NDI-Act-2023}, defines the governance-related parameters and clarifies which issuers are trusted and under what conditions credentials are revoked. Thus, Bhutan’s NDI exemplifies how SSI-aligned approaches are still dependent on centralised components, even on top of decentralised infrastructures.

In Buenos Aires, Argentina, QuarkID allows users to manage multiple categories of credentials, including civic records, licences and certificates \citep{QuarkID-Architecture}. It is also possible to self-issue credentials (albeit with limited recognition) \citep{QuarkID-Whitepaper-2022}. QuarkID’s infrastructure is anchored on zkSync and designed for interoperability with Ethereum-compatible networks such as Polygon and Rootstock, which define requirements for trust and co-ordination across multiple layers \citep{ToIP-DesignPrinciplesStack-2021}. Yet, despite its emphasis on decentralisation and alignment with the principles of SSI, QuarkID remains dependent on governmental authorities. This reveals that decentralised components often operate within constraints.

Japan’s My Number Card reflects a contrasting paradigm that is less decentralised by design but still relevant to user-centric approaches. It functions as an official identity document containing user attributes and a unique 12-digit identifier \citep{Japan-DigitalAgency-MyNumber-2025} that points to tax, social security or health records. A smartphone integration allows users to enrol on Apple iOS and Google Android devices \citep{Japan-DigitalAgency-MyNumber-Smartphone}. Despite the user-centric features, the overall approach remains dependent on a PKI-based trust anchor \citep{Japan-DigitalAgency-MyNumber-JPKI-Tools, Japan-DigitalAgency-JPKI-2025}. The smartphone integration also introduces dependencies on proprietary components \citep{Japan-DigitalAgency-MyNumber-Smartphone}, further reinforcing the centralisation. As such, Japan’s case emphasises that user-centric approaches do not necessarily involve decentralisation by default.

Although the above examples are not exhaustive, they illustrate a recurring pattern across real-world implementations. A consistent observation is that even the most decentralised architectures depend on forms of centralised governance or trust co-ordination. This again underscores the decentralisation paradox: user-centric approaches do not eliminate the involvement of authorities but rather redistribute and reframe it. In fact, governmental entities or other institutional actors continue to curate registries and trust lists that anchor trust \citep{Gans-2022, RamosFernandez-2024}. Thus, even as users obtain apparent control over their identity data, the surrounding infrastructure remains dependent on specific co-ordination mechanisms.

\section{Multi-Dimensional Analysis of Digital Identity (RQ2)}
\begin{figure}[!t]
\centering
\includegraphics[width=2in]{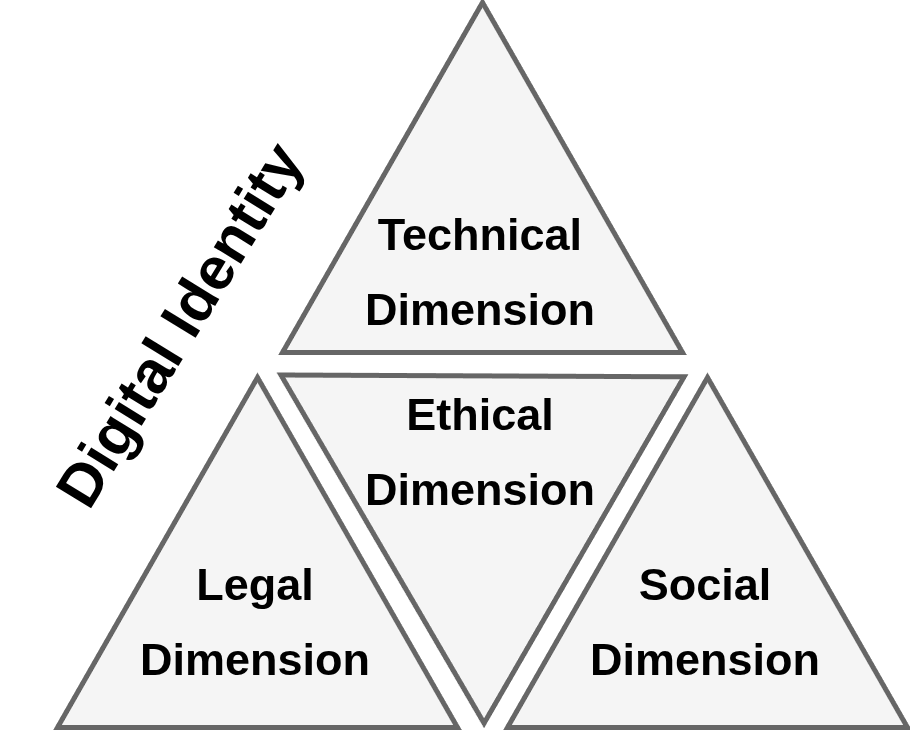}
\caption{The Digital Identity Tetrahedron: A conceptual model that illustrates the technical, legal, social and ethical dimensions of digital identity.}
\label{fig:digital-identity-tetrahedron}
\end{figure}

To understand the decentralisation paradox, digital identity is conceptualised through the Digital Identity Tetrahedron, which reflects its multidimensional nature (Figure \ref{fig:digital-identity-tetrahedron}). The four dimensions are derived from the classification of prior research presented in Table~\ref{tab:related-works-dimensions}, which illustrates how the selected literature addresses technical, legal, social and ethical aspects in a fragmented manner. The classification is interpretative and intended to illustrate the dominant thematic emphases in the literature rather than provide exhaustive or quantitative coding. Although additional dimensions could also be considered, these fall outside the scope of this paper. The tetrahedron therefore focuses on four dimensions that directly shape the decentralisation paradox:
\begin{enumerate}[label=(\alph*)]
  \item Technical Dimension \citep{Cucko-2023, Krul-2024, Schumm-2025}: It encompasses the architectural components, protocols, mechanisms and data structures that enable the identification, authentication, authorisation and lifecycle management of digital identities.
  \item Legal Dimension \citep{Andrasko-2021, Giannopoulou-2021, Podda-2025}: It relates to the principles, frameworks and legal requirements that define the recognition and governance of digital identities.
  \item Social Dimension \citep{Gans-2022, Lockl-2023, Teuschel-2023}: It reflects patterns of user behaviour, usability and accessibility, as well as the factors that influence adoption and digital inclusion or exclusion.
  \item Ethical Dimension \citep{Ishmaev-2021, Weigl-2023}: It concerns the moral and normative implications, including fairness, autonomy, power asymmetries and redistribution of responsibilities.
\end{enumerate}

\begin{table*}[t]
\centering
\caption{Classification of selected studies across technical, legal, social and ethical dimensions}
\label{tab:related-works-dimensions}
\begin{tabular}{lcccc}
\hline
Author(s) & Technical & Legal & Social & Ethical \\
\hline
\cite{Andrasko-2021}     &  & \cmark &  &  \\
\cite{Cucko-2023}        & \cmark &  &  &  \\
\cite{Degen-2024}        & \cmark & \cmark & \cmark &  \\
\cite{Gans-2022}         &  &  & \cmark &  \\
\cite{Ghaffari-2022}     & \cmark &  &  &  \\
\cite{Giannopoulou-2021} &  & \cmark & \cmark  &  \\
\cite{Ishmaev-2020}      &  &  &  & \cmark \\
\cite{Ishmaev-2021}      & \cmark &  & \cmark & \cmark \\
\cite{Kaplane-2025}      &  & \cmark & \cmark & \\
\cite{Kjorven-2026}      & \cmark & \cmark & \cmark & \\
\cite{Krul-2024}         & \cmark &  & \cmark & \\
\cite{Lockl-2023}        & \cmark & \cmark & \cmark &  \\
\cite{Martin-2024}       &  & \cmark & \cmark & \cmark \\
\cite{Podda-2025}        & \cmark & \cmark &  &  \\
\cite{Pohn-2024}         & \cmark &  & \cmark &  \\
\cite{Satybaldy-2024}    & \cmark & \cmark & \cmark  &  \\
\cite{Schardong-2022}    & \cmark & \cmark & \cmark &  \\
\cite{Schumm-2025}       & \cmark & \cmark & \cmark &  \\
\cite{Sharif-2025}       & \cmark &  &  &  \\
\cite{Teuschel-2023}     & \cmark &  & \cmark &  \\
\cite{Weigl-2023}        & \cmark  & \cmark & \cmark & \cmark \\
\hline
\end{tabular}
\end{table*}

\subsection{Technical Dimension}
From a technical point-of-view, user-centric approaches rely upon cryptographic mechanisms, standardised protocols and (sometimes) distributed ledgers. In essence, the specific design choices determine the degree of decentralisation. However, once the system boundaries are expanded to include the co-ordination components required for interoperability, governance and trust establishment, the decentralisation paradox emerges within the architecture. In this sense, decentralised presentation mechanisms depend on a series of recentralised technical dependencies that render such approaches trustworthy and operational in practice. The technical artefacts often involve:

\begin{itemize}
  \item Identity Data and Credential Structures: Decentralised Identifiers (DIDs) \citep{W3C-DID-2022} are independent of centralised registries and can be anchored on blockchains, distributed ledgers or peer-to-peer networks. In addition, Verifiable Credentials (VCs) \citep{W3C-VC-2025} and Selective Disclosure JSON Web Token (SD-JWT) VCs \citep{IETF-SD-JWT-VC} describe credentials that contain identity data in a privacy-preserving manner. For particular use-cases, there are even specifications such as mDLs \citep{ISO-18013-5-2021, ISO-18013-7-2025} and Digital Travel Credentials (DTCs) \citep{ICAO-DTC-PC-Protocols-2022, ICAO-DTC-GuidingCorePrinciples-2023}.

  \item Credential Exchange and Trust Mechanisms: There are multiple examples such as Decentralised Identifier Communication (DIDComm) \citep{DIF-DIDComm}, OpenID for Verifiable Credential Issuance (OID4VCI) \citep{OpenID4VCI-2025}, OpenID for Verifiable Presentation (OID4VP) \citep{OpenID4VP-2025} and Self-Issued OpenID Provider (SIOPv2) \citep{OIDC-SIOPv2-2023} that support the issuance, presentation and verification of credentials across architectures. These standardised protocols are often combined with trust registries (including PKI-based components) that define trusted issuers, verifiers and acceptable credential categories, thereby enabling the establishment of trust across organisational and jurisdictional boundaries.

  \item Digital Wallets: These are software (or, in some cases, hardware) solutions that serve as the user interfaces for managing credentials and cryptographic keys \citep{Babel-2025, Schumm-2025, Lukkien-2023}. Their design determines how users store, manage and present identity data. In practice, wallets need to be intuitive to use \citep{Supangkat-2025, Schumm-2025, Schardong-2022, Lukkien-2023}, resilient against malicious or unintended actions \citep{Soltani-2021, Pohn-2024, Sharif-2025, Ansaroudi-2025} and equipped with sufficient recovery methods \citep{Satybaldy-2024, Xian-2025}. Ideally, a well-designed digital wallet is meant to enable users to control their data with confidence. However, such protection often depends on hardware-backed cryptographic key management and platform-specific mechanisms, which introduce additional dependencies beyond the users’ control or awareness.
\end{itemize}

Beyond technical specifications, the success or failure of user-centric approaches depends on a series of legal and social enablers. On one hand, institutions and regulators shape the ecosystem through laws, trust frameworks and compliance requirements. On the other hand, human factors such as trust, usability and institutional acceptance are equally important \citep{Teuschel-2023}. Even if an architecture includes decentralised technical components, users could find it difficult to adopt or verifiers could refuse to accept credentials without a central guarantor. These considerations highlight that decentralised architectures do not operate in isolation, but rely on governance structures that define accountability, liability and trust relationships.

\subsection{Legal Dimension}
The question of governance was amongst the first to expose the tension between decentralisation and accountability in SSI \citep{Gans-2022, RamosFernandez-2024, Xian-2025}. For instance, the Sovrin network created its governance framework that detailed who could run network nodes, how issuers were on-boarded and how compliance with legislation was handled \citep{PavaDiaz-2024}. While not part of formal law, such frameworks often assume a quasi-legal character. In parallel, consortia defined governance models and sets of rules \citep{DIF-2025, ToIP-DesignPrinciplesStack-2021, ToIP-PrinciplesOfSSI-2021}. Hence, all these developments indicated that a trustless architecture cannot fulfil real-world requirements.

In practice, in service- or network-centric approaches, there are well-defined roles and responsibilities, e.g., governmental entities issue e-passports and financial institutions validate their customers' identities according to Know-Your-Customer (KYC) and Anti-Money Laundering (AML) legislation \citep{PavaDiaz-2024, Giannopoulou-2021}. However, such requirements become difficult to fulfil in user-centric approaches \citep{Sedlmeir-2021, Andrasko-2021}, as legal frameworks need to further detail the duties of issuers, users and verifiers. Consequently, decentralisation in digital identity faces pressure for accountability \citep{Xian-2025} which, in turn, prompts the introduction of centralised governance-related processes \citep{Gans-2022}; the decentralisation paradox reappears.

Furthermore, user-centric approaches might implicitly transform a series of technical assumptions into legal responsibilities that are, in turn, placed on users \citep{Kjorven-2026}. This raises critical questions regarding the allocation of liabilities, e.g., in cases of credential misuse, compromise or even revocation errors. Although multiple obligations shift to individuals, sole user control is often difficult to maintain in real-world environments due to structural power asymmetries between users and service providers \citep{Martin-2024}. This tension further illustrates the paradox, as the decentralisation of technical control is accompanied by additional governance and oversight mechanisms to address the emerging risks.

Last but not least, the legal notion of user control does not always address coercion or delegation \citep{Kaplane-2025}. In particular, users might be pressured into performing fraudulent transactions, disclosing excessive data or relying on intermediaries to access service providers. Consequently, legal frameworks often introduce relevant safeguards and oversight mechanisms to balance autonomy with accountability. In this sense, the decentralisation of technical control is accompanied by a recentralisation of legal responsibilities, illustrating another manifestation of the paradox.

\subsection{Social Dimension}
Besides the technical and legal dimensions, the adoption of user-centric approaches depends on whether users trust and accept them. The paradox is reflected in this dimension as well: centralisation might foster mistrust (e.g., fears of monopolisation or surveillance), while decentralisation might be perceived as fragmented or insecure \citep{Supangkat-2025, Ishmaev-2021, Kaplane-2025}. Meanwhile, the public often remains sceptical of digital identities and fears the potential erosion of civil liberties, regardless of the underlying architectural design and technical capabilities.

Moreover, there are persistent usability and inclusion-related challenges. Not all users are capable or willing to manage cryptographic keys or secure their own wallets \citep{Teuschel-2023, Lukkien-2023}. Therefore, several user-centric approaches incorporate social recovery mechanisms \citep{Satybaldy-2024, Xian-2025} which, in fact, re-introduce forms of delegated trust. In state-backed user-centric implementations, governmental entities might provide backup mechanisms, which could increase convenience but also recentralise control in case of inappropriate safeguards.

Another behavioural observation is that users often equate convenience with trust. The “privacy paradox” is also relevant in the context of digital identity \citep{Babel-2025, Lockl-2023, Teuschel-2023}, as users express concern about privacy and data protection but trade them for ease of use. Specifically, users might over-disclose personal data—without understanding the potential implications—if the interface makes transactions appear secure. Paradoxically, this could lead to greater data exposure than in service-centric or network-centric approaches, as digital wallets may create an unsubstantiated “illusion of control”. Thus, transparency, user education and responsible interface design are essential prerequisites \citep{Teuschel-2023, Xian-2025} to foster trust and adoption.

Ultimately, the value of digital identities depends on their recognition by SPs. Even if users maintain control over their credentials, their utilisation remains limited unless they are accepted across ecosystems. Moreover, user-centric approaches might increase cognitive burden \citep{Lockl-2023}, as users need to determine which credentials to share, with whom and under which conditions. These responsibilities might, in turn, lead to confusion and errors. Consequently, decentralisation still requires co-ordination mechanisms and trust arrangements, reinforcing another manifestation of the paradox.

\subsection{Ethical Dimension}
Is it ethical to shift a number of technical complexities to users (e.g., managing credentials and cryptographic keys), knowing that mistakes will happen? The ethical dimension revolves around the redistribution of responsibilities, mitigation of vulnerabilities and maintenance of control. As control moves to each user, so does the burden of managing risk. For example, the theft of a device (containing a digital wallet) could lead to the loss of access to essential data. Even mechanisms such as mnemonic-based or social (shared) recovery come with challenges, e.g., by requiring trusted contacts or third parties \citep{Satybaldy-2024, Schumm-2025, Xian-2025}. Therefore, the decentralisation paradox appears again, as the initial effort to eliminate intermediaries creates dependencies on additional entities.

Another ethical concern is digital inclusion \citep{Lukkien-2023}. Although some population groups lack smartphones or digital literacy, user-centric approaches often assume both access to and competence with digital solutions \citep{Kaplane-2025}. An ethical design is meant to require alternative options (e.g., card-based credentials, delegated access or support through community centres) to ensure participation from everyone, including people with cognitive impairments or disabilities \citep{Cucko-2023, Satybaldy-2024, Xian-2025}. Meanwhile, governmental entities and institutions need to refrain from treating people differently or denying access to services if they do not possess a digital wallet. Nonetheless, the existence of legal safeguards does not guarantee that indirect coercion or exclusion will not emerge.

Furthermore, there are persistent privacy and data protection-related risks \citep{Supangkat-2025, Kaplane-2025}. Even decentralised technical components could be repurposed to control users rather than empower them. For instance, a user-centric architecture on its own cannot guarantee protection against coercion or misuse, as malicious actors or intermediaries might still track, profile or exclude users \citep{Pohn-2024, Sharif-2025, Alvarez-2026}. Such risks undermine user trust and discourage adoption, while reflecting the social dimension of the paradox. Even the technical dimension alone cannot ensure ethical outcomes; when legal safeguards are weak or ambiguous, ethical considerations become the final safeguard, highlighting the interdependence among the technical, legal, social and ethical dimensions.

This is also where the phenomenon of "identity monopolies" \citep{Soltani-2021, Ishmaev-2021} resurfaces. In practice, decentralisation redistributes (but does not remove) the centres of power. This is evident when a limited number of technological solutions or wallet providers hold dominant positions within an ecosystem. As indicated earlier in this paper, certification requirements and interoperability constraints further reinforce such concentrations \citep{Degen-2024}, making it difficult for alternative approaches to develop. Consequently, open standards and protocols are often implemented in ways that concentrate influence among a small set of actors. Ethical design therefore requires a critical awareness of how decentralised approaches lead to different forms of centralisation and power asymmetries.

Overall, the ethical dimension underscores that decentralisation is not inherently virtuous. For instance, a digital identity ecosystem—in the hands of malicious actors or even abusive governments—could be repurposed to profile or discriminate against users. Likewise, a technical design may protect against surveillance by SPs, but there are residual risks in case the underlying technologies are subverted \citep{Sharif-2025}, e.g., in case of digital wallets with hidden capabilities. As such, an ethical design aims for a balance between autonomy, security, innovation and accountability.

\section{Discussion: Decentralisation Paradox in Digital Identity (RQ3)}
The analysis through the Digital Identity Tetrahedron reveals the decentralisation paradox. Even within user-centric approaches that incorporate decentralised components, a degree of recentralisation is expected. However, the excessive concentration or redistribution of power could undermine the core benefits of decentralisation. Hence, trust requires a careful calibration across all dimensions: users need to trust a solution prior to adoption and this necessitates sufficient safeguards. It is a delicate balance that requires a multidisciplinary collaboration; not only among technologists, but also jurists, sociologists and ethicists from the earliest stages of design.

\begin{figure}[!t]
\centering
\includegraphics[width=2.8in]{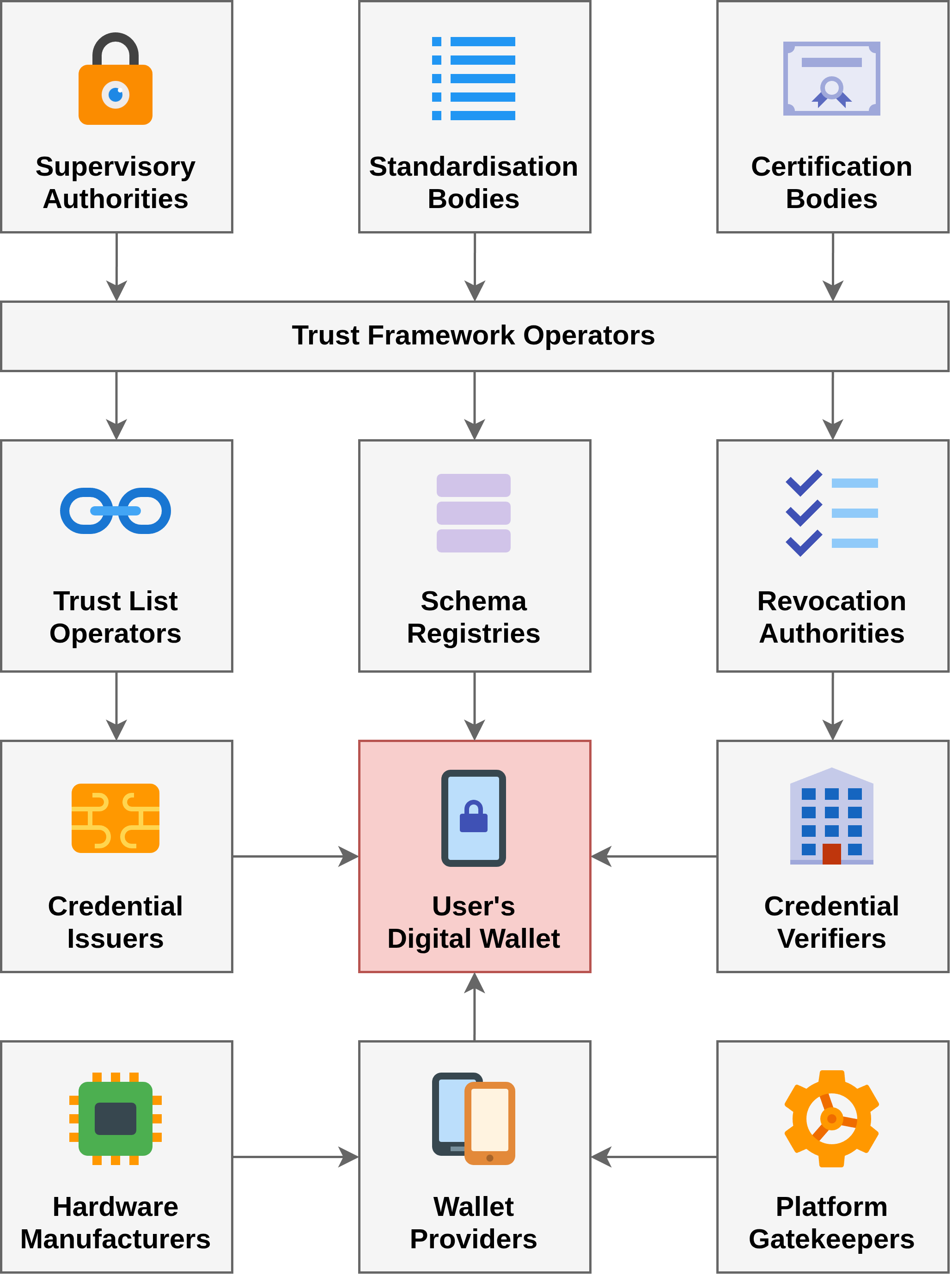}
\caption{The Decentralisation Paradox in Digital Identity: While user-centric approaches position digital wallets at the centre, their operation is shaped by multiple actors; this reveals reintroduced forms of centralisation. The arrows denote directional influence and asymmetric dependencies.}
\label{fig:decentralisation-paradox}
\end{figure}

Figure \ref{fig:decentralisation-paradox} illustrates the paradox. Despite the user's digital wallet being positioned at the centre, its operation depends on multiple actors. Specifically, supervisory authorities, standardisation and certification bodies formulate the trust framework, while trust list operators, schema registries and revocation authorities determine whether credentials and their presentations are valid. Meanwhile, issuers and verifiers control the issuance and acceptance of credentials, whereas wallet providers, hardware manufacturers and platform gatekeepers introduce additional constraints. Consequently, user-centric approaches do not eliminate centralisation but redistribute it.

\subsection{A Single Point of Convergence and Exposure}
As discussed earlier, digital wallets promise a unified point of convergence for personal data. This enhances convenience but also concentrates a lot of data on the user’s device. Technically, wallets risk becoming high-value targets, as a single compromise could expose multiple credentials at once \citep{Pohn-2024, Sharif-2025, Ziegler-2025}. In effect, the digital wallets' strength as an one-stop solution is also their greatest weakness and constitutes a single point of failure.

A key promise of digital wallets is data minimisation, which enables users to disclose only the minimum amount of data for specific purposes \citep{Ishmaev-2021, Podda-2025}. In theory, minimisation also limits profiling \citep{Ishmaev-2020} and restricts issuers or verifiers from tracking when and how credentials are used. However, as evidenced in real-world implementations, researchers highlight that certain design choices could still undermine unlinkability \citep{Baum-2024, Alvarez-2026, Podda-2025, Kaplane-2025}, e.g.,  through collusions between issuers and verifiers. Moreover, the aggregation of multiple credentials within a single wallet introduces additional privacy and data protection-related risks. As a consequence, this raises the spectre of different forms of profiling \citep{Sharif-2025}: a wallet that consolidates diverse aspects of a user’s identity could impose unprecedented impacts if misused. This is where the paradox resurfaces; user-centric approaches reduce reliance on siloed IdPs but simultaneously aggregate personal data within users' wallets and increase their exposure.

\subsection{A Single Authority that Anchors Trust}
In effect, in national and international implementations, governmental entities and institutions remain the ultimate trust anchors. The state authorities (and their designated entities) issue credentials that all other parties need to recognise. Moreover, user-centric approaches introduce further centralised dependencies; even though identity data exist on the users' devices, their security often depends on specific technical components such as Trusted Execution Environments (TEEs) or Secure Enclaves (SEs) \citep{Yildiz-2023, Soltani-2021, Ansaroudi-2025}. This again reinforces the decentralisation paradox: even when digital wallets empower users at the edge, they restrict them to centralised trust mechanisms and an oligopoly of actors for specific functionalities \citep{Ghaffari-2022, Lepore-2024}.

Overall, the legal, social and ethical implications of such arrangements are significant. When trust is anchored in government-backed infrastructures, authorities retain the power to withdraw, revoke or invalidate credentials. If a malicious actor decides to unilaterally invalidate a user’s credentials, it could disrupt that user’s access to multiple services and effectively cut them off from society. This illustrates how user-centric architectures could still reproduce structural power asymmetries; it also raises serious concerns about accountability, due process and democratic safeguards.

\subsection{A Forgotten User: Cognitive Overload and Exclusion}
Amid the excitement that surrounds the technical innovation of digital wallets, there is a risk of forgetting the users these solutions are meant to empower. The literature indicates that wallets often impose additional cognitive burdens and complexities that undermine their intended benefits \citep{Schardong-2022, Teuschel-2023, Pohn-2021}. In addition, the absence of transparent user interfaces could reflect an illusion of control \citep{Teuschel-2023} and fail to provide adequate protection. In fact, the inherent cognitive overload could render these solutions frustrating or even intimidating to the average user. As such, the decentralisation paradox highlights a deeper concern: ignoring these realities could increase digital inequalities.

An even starker challenge is digital exclusion \citep{Supangkat-2025, Kaplane-2025}. It cannot be assumed that everyone possesses a smartphone or the digital literacy to use one in a confident manner. If digital wallets become the default mode of identification and authentication, those who lack devices or skills could be left behind \citep{Kaplane-2025}. Even if alternative solutions are offered (e.g., physical wallet carriers), they might be treated as second-class users. Thus, user-centric approaches run the danger of becoming exclusionary by design. The forgotten user is the one for whom a wallet represents exclusion instead of empowerment.

\subsection{A Responsibility Shift Disguised as Empowerment}
As noted in this paper, the introduction of digital wallets imposes a number of obligations \citep{Krul-2024, Sharif-2025} that users may struggle to fulfil. While they are expected to safeguard their credentials and use them appropriately, the applicable legal frameworks might nevertheless presume a failure in their duty (e.g., in cases of fraud or compromise) \citep{Kjorven-2026}. In this sense, user empowerment would become an illusion: a series of responsibilities and liabilities are shifted downward onto users without adequate support or meaningful recourse.

The decentralisation paradox therefore raises broader ethical questions regarding the redistribution of such responsibilities and liabilities. Moreover, verifiers could pressure users into disclosing additional credentials and revealing more data than strictly necessary, thereby undermining data minimisation principles. In response, new intermediaries such as custodial wallet providers or managed identity services could emerge to assist users with handling their credentials \citep{Krul-2024, Cucko-2023}. However, such arrangements reintroduce reliance on third parties, ultimately undermining the autonomy that user-centric approaches promise.

\section{Conclusion}
The future of user-centric approaches depends on careful design choices at the intersection of the technical, legal, social and ethical dimensions. As these evolve, there is an urgent need to develop a richer conceptual framework for digital identity: one that focuses not only on architectural decentralisation, but also on who holds power, who absorbs risk and how such technological solutions ultimately serve the public good. The Digital Identity Tetrahedron provides one such lens, supporting the analysis of how decisions in one dimension could introduce unintended consequences in others.

A phenomenon that can be described as the "simplification paradox" represents a specific manifestation of the broader decentralisation paradox. Often, digital wallets are presented as instruments that optimise the exchange of identity data by consolidating credentials and interactions into a single interface. However, this apparent simplification at the user interface level conceals increased complexities in the underlying ecosystem. As responsibilities are redistributed across issuers, wallet providers, trust frameworks, certification bodies and platform operators, risks do not disappear but are shifted elsewhere. These hidden dependencies, which are analysable through the Digital Identity Tetrahedron, illustrate how simplification for users coincides with increased systemic complexities and fragmented data governance across actors.

These observations highlight the importance of holistic methodological approaches such as CST. Technical rigour alone might overlook that certain design choices introduce legal, social or ethical consequences elsewhere in the system. Without critical reflection, there is a risk of adopting architectures that are framed as user empowerment without substantiation. In essence, researchers and practitioners need to work across disciplines to bend the arc of implementation toward pragmatic user empowerment. This requires a careful balance between the promise of decentralisation and the practical realities of recentralisation. It is crucial to ensure that the next generation of digital identity fulfils its promises in practice rather than merely in rhetoric.

Furthermore, it is important to critically examine the meaning of user empowerment itself. Empowerment cannot be measured solely by the amount of identity data users control, but also by their ability to understand, manage and exercise that control in a meaningful manner. If empowerment depends on comprehension and confidence, then one-size-fits-all digital wallet solutions might not be sufficient. More adaptive, inclusive and supportive approaches are needed to accommodate diverse user capabilities, contexts and expectations. This suggests an important direction for future research on adaptive and user-centric digital identity approaches.

Finally, there are emerging developments such as Artificial Intelligence (AI)-driven agents which introduce additional complexities that remain unexplored in the current literature. Most approaches assume human users who present credentials and make decisions themselves. However, non-human agents acting on behalf of users can request services, present credentials and interact with verifiers. This raises important questions regarding liability, accountability and consent. For example, if an AI agent discloses credentials incorrectly, exceeds a mandate or performs actions that affect a user’s legal or financial standing, it becomes unclear who bears responsibility: the user, the wallet provider, the verifier or the agent’s developer? These scenarios extend the decentralisation and simplification paradoxes into a new domain, where autonomy is delegated to algorithmic mechanisms. A thorough understanding of these dynamics will be crucial for the next generation of user-centric digital identity approaches.

\bibliography{sn-bibliography}

@article{Cucko-2021,
  title   = {{Decentralized and Self-Sovereign Identity: Systematic Mapping Study}},
  author  = {{\v{S}}pela {{\v{C}}}u{\v{c}}ko and Muhamed Turkanovi\'c},
  journal = {IEEE Access},
  volume  = {9},
  pages   = {139009--139027},
  year    = {2021},
  url     = {https://doi.org/10.1109/access.2021.3117588}
}

@article{Ishmaev-2021,
  title   = {{Sovereignty, Privacy, and Ethics in Blockchain-Based Identity Management Systems}},
  author  = {Georgy Ishmaev},
  journal = {Ethics and Information Technology},
  volume  = {23},
  pages   = {239--252},
  year    = {2021},
  url     = {https://doi.org/10.1007/s10676-020-09563-x}
}

@article{Pohn-2021,
  title   = {{eID and Self-Sovereign Identity Usage: An Overview}},
  author  = {Daniela P\"ohn and Michael Grabatin and Wolfgang Hommel},
  journal = {Electronics},
  volume  = {10},
  number  = {22},
  pages   = {2811},
  year    = {2021},
  url     = {https://doi.org/10.3390/electronics10222811}
}

@article{Sedlmeir-2021,
  title   = {{Digital Identities and Verifiable Credentials}},
  author  = {Johannes Sedlmeir and Reilly Smethurst and Alexander Rieger and Gilbert Fridgen},
  journal = {Business \& Information Systems Engineering},
  volume  = {63},
  pages   = {603--613},
  year    = {2021},
  url     = {https://doi.org/10.1007/s12599-021-00722-y}
}

@article{Soltani-2021,
  title   = {{A Survey of Self-Sovereign Identity Ecosystem}},
  author  = {Reza Soltani and Uyen Trang Nguyen and Aijun An},
  journal = {Security and Communication Networks},
  volume  = {2021},
  year    = {2021},
  pages   = {8873429},
  url     = {https://doi.org/10.1155/2021/8873429}
}

@article{Alizadeh-2022,
  title   = {{Comparative Analysis of Decentralized Identity Approaches}},
  author  = {Morteza Alizadeh and Karl Andersson and Olov Schel\'en},
  journal = {IEEE Access},
  volume  = {10},
  pages   = {92273--92283},
  year    = {2022},
  url     = {https://doi.org/10.1109/access.2022.3202553}
}

@inproceedings{Lips-2022,
  title     = {{Re-Shaping the EU Digital Identity Framework}},
  author    = {Silvia Lips and Natalia Vinogradova and Robert Krimmer and Dirk Draheim},
  booktitle = {Proceedings of the 23rd Annual International Conference on Digital Government Research},
  publisher = {Association for Computing Machinery},
  year      = {2022},
  pages     = {13--21},
  url       = {https://doi.org/10.1145/3543434.3543652}
}

@article{Schardong-2022,
  author  = {Frederico Schardong and Ricardo Cust{\'o}dio},
  title   = {Self-Sovereign Identity: A Systematic Review, Mapping and Taxonomy},
  journal = {Sensors},
  volume  = {22},
  number  = {15},
  pages   = {5641},
  year    = {2022},
  url     = {https://doi.org/10.3390/s22155641},
}

@inproceedings{Lukkien-2023,
  title     = {{Barriers for Developing and Launching Digital Identity Wallets}},
  author    = {Bert Lukkien and Nitesh Bharosa and Mark De Reuver},
  booktitle = {Proceedings of the 24th Annual International Conference on Digital Government Research},
  publisher = {Association for Computing Machinery},
  year      = {2023},
  pages     = {289--299},
  url       = {https://doi.org/10.1145/3598469.3598501}
}

@article{Tan-2023,
  author    = {Kheng Leong Tan and Chi-Hung Chi and Kwok-Yan Lam},
  title     = {{Survey on Digital Sovereignty and Identity: From Digitization to Digitalization}},
  journal   = {ACM Computing Surveys},
  volume    = {56},
  number    = {3},
  pages     = {61},
  year      = {2023},
  publisher = {Association for Computing Machinery},
  url       = {https://doi.org/10.1145/3616400},
}

@article{Degen-2024,
  title   = {{Wallet Wars or Digital Public Infrastructure? Orchestrating a Digital Identity Data Ecosystem from a Government Perspective}},
  author  = {Konrad Degen and Timm Teubner},
  journal = {Electronic Markets},
  volume  = {34},
  pages   = {50},
  year    = {2024},
  url     = {https://doi.org/10.1007/s12525-024-00731-1}
}

@article{Krul-2024,
  title   = {{SoK: Trusting Self-Sovereign Identity}},
  author  = {Evan Krul and Hye-young Paik and Sushmita Ruj and Salil S. Kanhere},
  journal = {Proceedings on Privacy Enhancing Technologies},
  volume  = {2024},
  number  = {3},
  pages   = {297--313},
  year    = {2024},
  url     = {https://doi.org/10.56553/popets-2024-0079}
}

@inproceedings{Lepore-2024,
  title     = {{Aligning eIDAS and Trust Over IP: A Mapping Approach}},
  author    = {Cristian Lepore and Romain Laborde and Jessica Eynard},
  booktitle = {Proceedings of the 19th International Conference on Availability, Reliability and Security},
  publisher = {Association for Computing Machinery},
  year      = {2024},
  pages     = {Article 186},
  url       = {https://doi.org/10.1145/3664476.3670919}
}

@Article{Pohn-2024,
  author    = {Daniela Pöhn and Michael Grabatin and Wolfgang Hommel},
  title     = {{Analyzing the Threats to Blockchain-Based Self-Sovereign Identities by Conducting a Literature Survey}},
  journal   = {Applied Sciences},
  volume    = {14},
  number    = {1},
  article-number = {139},
  year      = {2024},
  issn      = {2076-3417},
  url       = {https://doi.org/10.3390/app14010139}
}

@article{RamosFernandez-2024,
  title   = {{Evaluation of Trust Service and Software Product Regimes for {{Zero-Knowledge}} Proof Development under {{eIDAS}} 2.0}},
  author  = {Ra\"ul {Ramos Fern\'andez}},
  journal = {Computer Law \& Security Review},
  volume  = {53},
  pages   = {105968},
  year    = {2024},
  url     = {https://doi.org/10.1016/j.clsr.2024.105968}
}

@Article{Satybaldy-2024,
  author    = {Abylay Satybaldy and Mohammad Sadek Ferdous and Mariusz Nowostawski},
  title     = {{A Taxonomy of Challenges for Self-Sovereign Identity Systems}},
  journal   = {IEEE Access},
  volume    = {12},
  pages     = {16151--16177},
  year      = {2024},
  url       = {https://doi.org/10.1109/access.2024.3357940}
}

@InProceedings{Sharif-2025,
  author    = {Amir Sharif and Zahra Ebadi Ansaroudi and Giada Sciarretta and Daniela P{\"o}hn and Majid Mollaeefar and Wolfgang Hommel and Silvio Ranise},
  editor    = {Simon Collart-Dutilleul and Samir Ouchani and Nora Cuppens and Fr{\'e}d{\'e}ric Cuppens},
  title     = {{Protecting Digital Identity Wallet: A Threat Model in the Age of eIDAS 2.0}},
  booktitle = {Risks and Security of Internet and Systems},
  year      = {2025},
  publisher = {Springer Nature Switzerland},
  pages     = {89--106},
  url       = {https://doi.org/10.1007/978-3-031-89350-6_6}
}

@article{Ansaroudi-2025,
  title   = {{Navigating Secure Storage Requirements for {{EUDI}} Wallets: A Review Paper}},
  author  = {Zahra Ebadi Ansaroudi and Giada Sciarretta and Andrea De Maria and Silvio Ranise},
  journal = {EURASIP Journal on Information Security},
  volume  = {2025},
  number  = {2},
  pages   = {1--24},
  year    = {2025},
  url     = {https://doi.org/10.1186/s13635-025-00187-6}
}

@article{Babel-2025,
  title   = {{Self-Sovereign Identity and Digital Wallets}},
  author  = {Matthias Babel and Lukas Willburger and Jonathan Lautenschlager and Fabiane V\"olter and Tobias Guggenberger and Marc-Fabian K\"orner and Johannes Sedlmeir and Jens Str\"uker and Nils Urbach},
  journal = {Electronic Markets},
  volume  = {35},
  pages   = {28},
  year    = {2025},
  url     = {https://doi.org/10.1007/s12525-025-00772-0}
}

@Article{Ziegler-2025,
  author    = {Leonhard Ziegler and Michael Grabatin and Daniela P{\"o}hn and Wolfgang Hommel},
  title     = {{Designing a Security Incident Response Process for Self-Sovereign Identities}},
  journal   = {EURASIP Journal on Information Security},
  volume    = {2025},
  number    = {12},
  year      = {2025},
  url       = {https://doi.org/10.1186/s13635-025-00195-6},
}

@Article{Supangkat-2025,
  author    = {Suhono Harso Supangkat and Hendra Sandhi Firmansyah and Irma Rizkia and Rezky Kinanda},
  title     = {{Challenges in Implementing Cross-Border Digital Identity Systems for Global Public Infrastructure: A Comprehensive Analysis}},
  journal   = {IEEE Access},
  volume    = {13},
  pages     = {42083--42098},
  year      = {2025},
  url       = {https://doi.org/10.1109/access.2025.3547373}
}

@article{Schumm-2025,
  author    = {Daria Schumm and Katharina O. E. Müller and Burkhard Stiller},
  title     = {Are We There Yet? A Study of Decentralized Identity Applications},
  journal   = {IEEE Access},
  year      = {2025},
  volume    = {13},
  pages     = {125232--125259},
  url       = {https://doi.org/10.1109/access.2025.3588170}
}

@Article{Lockl-2023,
  author    = {Jannik Lockl and Nico Thanner and Manuel Utz and Maximilian R{\"o}glinger},
  title     = {{The Paradoxical Impact of Information Privacy on Privacy Preserving Technology: The Case of Self-Sovereign Identities}},
  journal   = {International Journal of Innovation and Technology Management},
  volume    = {20},
  number    = {04},
  pages     = {2350025},
  year      = {2023},
  url       = {https://doi.org/10.1142/S0219877023500256}
}

@Article{Martin-2024,
  author    = {Nicholas Martin and Frederik M. Metzger},
  title     = {{The Chimera of Control: Self-Sovereign Identity, Data Control, and User Perceptions}},
  journal   = {Human Technology},
  volume    = {20},
  number    = {2},
  pages     = {183--223},
  year      = {2024},
  url       = {https://doi.org/10.14254/1795-6889.2024.20-2.1}
}

@Article{Manimaran-2025,
  author    = {Praveensankar Manimaran and Thiago Garrett and Leander Jehl and Roman Vitenberg},
  title     = {{Decentralization Trends in Identity Management: From Federated to Self-Sovereign Identity Management Systems}},
  journal   = {Computer Science Review},
  volume    = {58},
  pages     = {100776},
  year      = {2025},
  url       = {https://doi.org/10.1016/j.cosrev.2025.100776}
}

@article{Weigl-2023,
  title   = {{The Construction of Self-Sovereign Identity: Extending the Interpretive Flexibility of Technology Towards Institutions}},
  author  = {Linda Weigl and Tom Barbereau and Gilbert Fridgen},
  journal = {Government Information Quarterly},
  volume  = {40},
  number  = {4},
  pages   = {101873},
  year    = {2023},
  url     = {https://doi.org/10.1016/j.giq.2023.101873}
}

@misc{ARF-2025-Ecosystem,
  title  = {{EUDI Wallet Ecosystem -- European Digital Identity Wallet Architecture and Reference Framework v2.8.0}},
  author = {{European Commission}},
  year   = {2026},
  url    = {https://eudi.dev/2.8.0/architecture-and-reference-framework-main/#3-eudi-wallet-ecosystem}
}

@misc{ARF-2025-TrustModel,
  title  = {{Trust Model -- European Digital Identity Wallet Architecture and Reference Framework v2.8.0}},
  author = {{European Commission}},
  year   = {2026},
  url    = {https://eudi.dev/2.8.0/architecture-and-reference-framework-main/#6-trust-model}
}

@misc{ARF-2025-Certification,
  title  = {{Certification and Risk Management -- European Digital Identity Wallet Architecture and Reference Framework v2.8.0}},
  author = {{European Commission}},
  year   = {2026},
  url    = {https://eudi.dev/2.8.0/architecture-and-reference-framework-main/#7-certification-and-risk-management}
}

@misc{NIST-2024-SP800-63,
  title  = {{NIST SP 800-63: Digital Identity Guidelines}},
  author = {{National Institute of Standards and Technology (NIST)}},
  year   = {2024},
  url    = {https://pages.nist.gov/800-63-4/sp800-63.html}
}

@misc{NIST-2024-SP800-63A,
  title  = {{NIST SP 800-63A: Enrollment and Identity Proofing}},
  author = {{National Institute of Standards and Technology (NIST)}},
  year   = {2024},
  url    = {https://pages.nist.gov/800-63-4/sp800-63a.html}
}

@misc{EU-2024-1183,
  title  = {{Regulation (EU) 2024/1183 of the European Parliament and of the Council of 11 April 2024 Amending Regulation (EU) No 910/2014 as Regards Establishing the European Digital Identity Framework}},
  author = {{European Parliament and Council of the European Union}},
  year   = {2024},
  url    = {https://eur-lex.europa.eu/eli/reg/2024/1183/oj}
}

@misc{W3C-VC-2025,
  title = {{Verifiable Credentials Data Model v2.0}},
  author = {{World Wide Web Consortium (W3C)}},
  year = {2025},
  url = {https://www.w3.org/TR/vc-data-model-2.0/}
}

@misc{Arizona-mDL,
  title        = {{Arizona Mobile Driver License (mDL)}},
  author       = {{Arizona Department of Transportation}},
  year         = {2026},
  url          = {https://azdot.gov/mvd/mobile-driver-license}
}

@misc{California-mDL,
  title        = {{California Mobile Driver License (mDL)}},
  author       = {{California Department of Motor Vehicles}},
  year         = {2026},
  url          = {https://www.dmv.ca.gov/portal/california-mdl/}
}

@misc{Georgia-mDL,
  title        = {{Georgia Digital Driver's License and ID}},
  author       = {{Georgia Department of Driver Services}},
  year         = {2026},
  url          = {https://dds.georgia.gov/georgia-licenseid/ga-digital-id}
}

@misc{GoogleWallet-2025,
  title        = {{Store Your Digital ID on Your Phone -- Google Wallet}},
  author       = {Google},
  year         = {2026},
  url          = {https://wallet.google/intl/en_us/digitalid/}
}

@misc{W3C-DID-2022,
  title = {{Decentralized Identifiers (DIDs) v1.0: Core Architecture, Data Model, and Representations}},
  author = {{World Wide Web Consortium (W3C)}},
  year = {2022},
  url = {https://www.w3.org/TR/did-1.0/}
}

@misc{OpenID4VCI-2025,
  title = {{OpenID for Verifiable Credential Issuance 1.0}},
  author = {{OpenID Foundation}},
  year = {2025},
  url = {https://openid.net/specs/openid-4-verifiable-credential-issuance-1_0.html}
}

@misc{OpenID4VP-2025,
  title = {{OpenID for Verifiable Presentations 1.0}},
  author = {{OpenID Foundation}},
  year = {2025},
  url = {https://openid.net/specs/openid-4-verifiable-presentations-1_0.html}
}

@misc{DIF-DIDComm,
  author       = {{Decentralized Identity Foundation (DIF)}},
  title        = {{DIDComm Messaging v2.1}},
  year         = {2023},
  url          = {https://identity.foundation/didcomm-messaging/spec/v2.1}
}

@misc{IETF-SD-JWT-VC,
  title        = {{Selective Disclosure for JWT-Based Verifiable Credentials (SD-JWT VC)}},
  author       = {{Internet Engineering Task Force (IETF)}},
  year         = {2026},
  url          = {https://datatracker.ietf.org/doc/draft-ietf-oauth-sd-jwt-vc/}
}

@misc{ISO-18013-5-2021,
  author       = {{International Organization for Standardization (ISO)}},
  title        = {{ISO/IEC 18013-5:2021 -- Personal Identification: ISO-Compliant Driving Licence -- Part 5: Mobile Driving Licence (mDL) Application}},
  year         = {2021},
  url          = {https://www.iso.org/standard/69084.html}
}

@misc{ISO-18013-7-2025,
  author       = {{International Organization for Standardization (ISO)}},
  title        = {{ISO/IEC TS 18013-7:2025 -- Personal Identification: ISO-Compliant Driving Licence -- Part 7: Mobile Driving Licence (mDL) Add-On Functions}},
  year         = {2025},
  url          = {https://www.iso.org/standard/91154.html}
}

@misc{OIDC-SIOPv2-2023,
  title = {{Self-Issued OpenID Provider v2}},
  author = {{OpenID Foundation}},
  year = {2023},
  url = {https://openid.net/specs/openid-connect-self-issued-v2-1_0.html}
}

@misc{Apple-Wallet-ID-States,
  title        = {{ID in Wallet –- Supported U.S. States}},
  author       = {{Apple}},
  year         = {2025},
  url          = {https://learn.wallet.apple/id#states-list}
}

@misc{ICAO-DTC-GuidingCorePrinciples-2023,
  title        = {{Machine Readable Travel Documents Technical Report: Guiding Core Principles for the Development of Digital Travel Credential (DTC)}},
  author       = {{International Civil Aviation Organization (ICAO)}},
  year         = {2023},
  url          = {https://www2023.icao.int/Security/FAL/TRIP/PublishingImages/Pages/Publications/ICAO%20GUIDING%20CORE%20PRINCIPLES%20DTC_Draft%20v4.8.pdf}
}

@misc{ICAO-DTC-PC-Protocols-2022,
  title        = {{Machine Readable Travel Documents Technical Report: Digital Travel Credentials (DTC) Physical Component and Protocols}},
  author       = {{International Civil Aviation Organization (ICAO)}},
  year         = {2022},
  url          = {https://www2023.icao.int/Security/FAL/TRIP/PublishingImages/Pages/Publications/ICAO%20TR%20-%20Digital%20Travel%20Credentials%20PC.pdf}
}

@misc{Japan-DigitalAgency-MyNumber-2025,
  title        = {{My Number (Individual Number) Scheme / My Number Card}},
  author       = {{Digital Agency, Government of Japan}},
  year         = {2025},
  url          = {https://www.digital.go.jp/en/policies/mynumber}
}

@misc{Japan-DigitalAgency-JPKI-2025,
  title        = {{Japanese Public Key Infrastructure (JPKI)}},
  author       = {{Digital Agency, Government of Japan}},
  year         = {2026},
  url          = {https://www.digital.go.jp/en/policies/mynumber/private-business/jpki-introduction}
}

@misc{Japan-DigitalAgency-MyNumber-JPKI-Tools,
  title        = {{Common Infrastructure and General-Purpose Tools for My Number Card Use}},
  author       = {{Digital Agency, Government of Japan}},
  year         = {2024},
  url          = {https://www.digital.go.jp/en/policies/mynumber/local-government/platforms-and-generic-tools-for-jpki}
}

@misc{Japan-DigitalAgency-MyNumber-Smartphone,
  title        = {{Smartphone My Number Card}},
  author       = {{Digital Agency, Government of Japan}},
  year         = {2026},
  url          = {https://www.digital.go.jp/en/policies/mynumber/smartphone-certification}
}

@misc{QuarkID-Whitepaper-2022,
  title        = {{Self-Sovereign Identity: Basis of a New Decentralized Digital Ecosystem (QuarkID Whitepaper)}},
  author       = {{QuarkID}},
  year         = {2022},
  url          = {https://github.com/ssi-quarkid/WhitePaper}
}

@misc{QuarkID-Architecture,
  title        = {{QuarkID Architecture}},
  author       = {{QuarkID}},
  year         = {2025},
  url          = {https://docs.quarkid.org/en/Arquitectura/arquitectura/}
}

@misc{ToIP-BhutanNDI-CaseStudy-2024,
  title        = {{Bhutan NDI (National Digital Identity) \& ToIP Digital Trust Ecosystems}},
  author       = {{Trust over IP (ToIP) Foundation}},
  year         = {2024},
  url          = {https://trustoverip.org/wp-content/uploads/Case-Study-Bhutan-NDI-National-Digital-Identity-ToIP-Digital-Trust-Ecosystems-V1.0-2024-05-21.ext_.pdf}
}

@misc{UNDP-Bhutan-DigitalStrategy-2024,
  title        = {{Kingdom of Bhutan: Digital Development and Transformation Strategy}},
  author       = {{United Nations Development Programme (UNDP)}},
  year         = {2024},
  url          = {https://www.undp.org/bhutan/publications/kingdom-bhutan-digital-development-and-transformation-strategy}
}

@misc{ISOIEC-24760-1-2025,
  title        = {{ISO/IEC 24760-1:2025 -- Information Security, Cybersecurity and Privacy Protection -- A Framework for Identity Management}},
  author       = {{International Organization for Standardization (ISO)}},
  year         = {2025},
  url          = {https://www.iso.org/standard/24760-1}
}

@misc{ITU-T-X1252-2021,
  title        = {{ITU-T Recommendation X.1252 -- Baseline Identity Management Terms and Definitions}},
  author       = {{International Telecommunication Union (ITU)}},
  year         = {2021},
  url          = {https://www.itu.int/rec/T-REC-X.1252-202104-I/en}
}

@misc{ToIP-PrinciplesOfSSI-2021,
  title        = {{Principles of SSI}},
  author       = {{Trust Over IP Foundation (ToIP)}},
  year         = {2021},
  url          = {https://trustoverip.org/wp-content/uploads/2021/10/ToIP-Principles-of-SSI.pdf}
}

@misc{ToIP-DesignPrinciplesStack-2021,
  title        = {{Design Principles for the Trust over IP Stack}},
  author       = {{Trust Over IP Foundation (ToIP)}},
  year         = {2021},
  url          = {https://trustoverip.org/wp-content/uploads/Design-Principles-for-the-ToIP-Stack-V1.0-2022-11-17.pdf}
}

@article{Xian-2025,
  title   = {{A Survey on Decentralized Identity Management Systems}},
  author  = {Jia Xian and Lin You and Qifei Yi and Jixiang Wang and Gengran Hu},
  journal = {Computer Science Review},
  volume  = {58},
  pages   = {100811},
  year    = {2025},
  url     = {https://doi.org/10.1016/j.cosrev.2025.100811}
}

@article{Alvarez-2026,
  title   = {{Privacy Evaluation of the European Digital Identity Wallet's Architecture and Reference Framework}},
  author  = {Ivan Abellan Alvarez and Pol Holzmer and Johannes Sedlmeir},
  journal = {Computers \& Security},
  volume  = {160},
  pages   = {104707},
  year    = {2026},
  url     = {https://doi.org/10.1016/j.cose.2025.104707}
}

@article{Kaplane-2025,
  title   = {{The European Digital Identity Wallet: A New Human Right Unlocked?}},
  author  = {Anastasija Kaplane},
  journal = {Nordic Journal of Human Rights},
  volume  = {43},
  number  = {3},
  pages   = {304--316},
  year    = {2025},
  url     = {https://doi.org/10.1080/18918131.2025.2551458}
}

@article{Podda-2025,
  title   = {{The Impact of Zero-Knowledge Proofs on Data Minimisation Compliance of Digital Identity Wallets}},
  author  = {Emanuela Podda and Pol Holzmer and Alexandre Amard and Johannes Sedlmeir and Gilbert Fridgen},
  journal = {Internet Policy Review},
  volume  = {14},
  number  = {3},
  pages   = {1--29},
  year    = {2025},
  url     = {https://doi.org/10.14763/2025.3.2019}
}

@article{PavaDiaz-2024,
  title   = {{Self-Sovereign Identity on the Blockchain: Contextual Analysis and Quantification of SSI Principles Implementation}},
  author  = {Roberto A. Pava-Diaz and Jesus Gil-Ruiz and Danilo A. Lopez-Sarmiento},
  journal = {Frontiers in Blockchain},
  volume  = {7},
  pages   = {1443362},
  year    = {2024},
  url     = {https://doi.org/10.3389/fbloc.2024.1443362}
}

@article{Cucko-2023,
  title   = {{Towards a Catalogue of Self-Sovereign Identity Design Patterns}},
  author  = {{\v{S}}pela {{\v{C}}}u{\v{c}}ko and Vid Ker{\v{s}}i{\v{c}} and Muhamed Turkanovi\'c},
  journal = {Applied Sciences},
  volume  = {13},
  number  = {9},
  pages   = {5395},
  year    = {2023},
  url     = {https://doi.org/10.3390/app13095395}
}

@article{Teuschel-2023,
  title   = {{Don't Annoy Me With Privacy Decisions! Designing Privacy-Preserving User Interfaces for SSI Wallets on Smartphones}},
  author  = {Moritz Teuschel and Daniela P{\"o}hn and Michael Grabatin and Felix Dietz and Wolfgang Hommel and Florian Alt},
  journal = {IEEE Access},
  volume  = {11},
  pages   = {131814--131835},
  year    = {2023},
  url     = {https://doi.org/10.1109/ACCESS.2023.3334908}
}

@article{Yildiz-2023,
  title   = {{Toward Interoperable Self-Sovereign Identities}},
  author  = {Hakan Yildiz and Axel K{\"u}pper and Dirk Thatmann and Sebastian Gondor and Patrick Herbke},
  journal = {IEEE Access},
  volume  = {11},
  pages   = {114080--114116},
  year    = {2023},
  url     = {https://doi.org/10.1109/ACCESS.2023.3313723}
}

@article{Gans-2022,
  title   = {{Governance and Societal Impact of Blockchain-Based Self-Sovereign Identities}},
  author  = {Rachel Benchaya Gans and Jolien Ubacht and Marijn Janssen},
  journal = {Policy and Society},
  volume  = {41},
  number  = {3},
  pages   = {402--413},
  year    = {2022},
  url     = {https://doi.org/10.1093/polsoc/puac018}
}

@article{Cucko-2022,
  title   = {{Towards the Classification of Self-Sovereign Identity Properties}},
  author  = {{\v{S}}pela {{\v{C}}}u{\v{c}}ko and {\v{S}}eila Be{\'c}irovi{\'c} and Aida Kamisalic and Sa{\v{s}}a Mrdovic and Muhamed Turkanovi{\'c}},
  journal = {IEEE Access},
  volume  = {10},
  pages   = {88306--88329},
  year    = {2022},
  url     = {https://doi.org/10.1109/ACCESS.2022.3199414}
}

@article{Andrasko-2021,
  title   = {{Those Who Shall be Identified: The Data Protection Aspects of the Legal Framework for Electronic Identification in the European Union}},
  author  = {Jozef Andrasko and Matus Mesarcik},
  journal = {TalTech Journal of European Studies},
  volume  = {11},
  number  = {2},
  pages   = {3--24},
  year    = {2021},
  url     = {https://doi.org/10.2478/bjes-2021-0012}
}

@article{Ghaffari-2022,
  title   = {{Identity and Access Management using Distributed Ledger Technology: A Survey}},
  author  = {Fariba Ghaffari and Komal Gilani and Emmanuel Bertin and Noel Crespi},
  journal = {International Journal of Network Management},
  volume  = {32},
  number  = {2},
  pages   = {e2180},
  year    = {2022},
  url     = {https://doi.org/10.1002/nem.2180}
}

@article{Giannopoulou-2021,
  title   = {{Self-Sovereign Identity}},
  author  = {Alexandra Giannopoulou and Fennie Wang},
  journal = {Internet Policy Review},
  volume  = {10},
  number  = {2},
  year    = {2021},
  url     = {https://doi.org/10.14763/2021.2.1550}
}

@article{Ishmaev-2020,
  title   = {{Identity Management Systems: Singular Identities and Multiple Moral Issues}},
  author  = {Georgy Ishmaev and Quinten Stokkink},
  journal = {Frontiers in Blockchain},
  volume  = {3},
  pages   = {15},
  year    = {2020},
  url     = {https://doi.org/10.3389/fbloc.2020.00015}
}

@book{Jackson-2024,
  author    = {Michael C. Jackson},
  title     = {{Critical Systems Thinking: A Practitioner's Guide}},
  year      = {2024},
  publisher = {Wiley},
  doi       = {https://doi.org/10.1002/9781394203604}
}

@book{Jackson-2019,
  author    = {Michael C. Jackson},
  title     = {{Critical Systems Thinking and the Management of Complexity}},
  year      = {2019},
  publisher = {Wiley},
}

@misc{Baum-2024,
  author       = {Carsten Baum and Olivier Blazy and Jan Camenisch and Jaap{-}Henk Hoepman and Eysa Lee and Anja Lehmann and Anna Lysyanskaya and Ren{\'e} Mayrhofer and Hart Montgomery and Ngoc Khanh Nguyen and Bart Preneel and Abhi Shelat and Daniel Slamanig and Stefano Tessaro and S{\o}ren Eller Thomsen and Carmela Troncoso},
  title        = {{Cryptographers' Feedback on the EU Digital Identity's ARF}},
  year         = {2024},
  url          = {https://github.com/eu-digital-identity-wallet/eudi-doc-architecture-and-reference-framework/issues/200}
}

@misc{Bhutan-NDI-Act-2023,
  author       = {{Parliament of Bhutan}},
  title        = {{National Digital Identity Act of Bhutan 2023}},
  year         = {2023},
  url          = {https://tech.gov.bt/wp-content/uploads/2024/09/National-Digital-Identity-Act-of-Bhutan-2023.pdf}
}

@misc{EU-2015-1502,
  title  = {{Commission Implementing Regulation (EU) 2015/1502 of 8 September 2015 Setting Out Minimum Technical Specifications and Procedures for Assurance Levels for Electronic Identification Means Pursuant to Article 8(3) of Regulation (EU) No 910/2014}},
  author = {{European Commission}},
  year   = {2015},
  url    = {https://eur-lex.europa.eu/eli/reg_impl/2015/1502/oj/eng}
}

@misc{DIF-2025,
  title        = {{About Decentralized Identity Foundation -- Governing Documents}},
  author       = {{Decentralized Identity Foundation}},
  year         = {2025},
  url          = {https://identity.foundation/governance/about},
  note         = {Accessed: 2025-11-25}
}

@article{Konstantinidis-2026,
  title   = {{The Paradigm Shift Toward User-Centric Digital Identity Approaches: A Systematic Literature Review}},
  author  = {Ioannis Konstantinidis and Ioannis Mavridis and Evangelos K. Markakis},
  journal = {IEEE Access},
  volume  = {14},
  pages   = {58901--58918},
  year    = {2026},
  url     = {https://doi.org/10.1109/ACCESS.2026.3680835}
}

@article{Kjorven-2026,
  title   = {{Safe and Inclusive or Unsafe and Discriminatory? European Digital Identity Wallets and the Challenges of ‘Sole Control’}},
  author  = {Marte Eidsand Kj{\o}rven and Kristian Gj{\o}steen and Tone Linn W{\ae}rstad},
  journal = {Computer Law \& Security Review},
  volume  = {60},
  pages   = {106235},
  year    = {2026},
  doi     = {10.1016/j.clsr.2025.106235}
}

\end{document}